\documentclass[aps,prb,twocolumn,showpacs,showkeys,groupaddress]{revtex4}

\usepackage{amsmath}
\usepackage{amssymb}
\usepackage{epsfig}
\usepackage{bm}
\usepackage{longtable}
\usepackage{graphicx}
\usepackage{graphics}

\begin{document}

\title{
Imaging correlated wave functions of few-electron quantum dots: \\
Theory and scanning tunneling spectroscopy experiments
}

\author{Massimo Rontani}
\email{rontani@unimore.it}
\homepage{http://www.nanoscience.unimo.it/max_index.html}
\affiliation{CNR-INFM National Research Center S3 
and Dipartimento di Fisica, Universit\`a degli Studi
di Modena e Reggio Emilia,
Via Campi 213/A, 41100 Modena, Italy}

\author{Elisa Molinari}
\affiliation{CNR-INFM National Research Center S3
and Dipartimento di Fisica, Universit\`a degli Studi
di Modena e Reggio Emilia,
Via Campi 213/A, 41100 Modena, Italy}

\author{Giuseppe Maruccio}
\thanks{Permanent address: CNR-INFM National Research Center NNL,
Distretto Tecnologico ISUFI Via Arnesano, 73100 Lecce, 
Italy}
\email{giuseppe.maruccio@unile.it}
\affiliation{Institute of Applied Physics, University of Hamburg,
Jungiusstrasse 11 20355 Hamburg, Germany}

\author{Martin Janson}
\affiliation{Institute of Applied Physics, University of Hamburg,
Jungiusstrasse 11 20355 Hamburg, Germany}

\author{Andreas Schramm}
\affiliation{Institute of Applied Physics, University of Hamburg,
Jungiusstrasse 11 20355 Hamburg, Germany}

\author{Christian Meyer}
\affiliation{Institute of Applied Physics, University of Hamburg,
Jungiusstrasse 11 20355 Hamburg, Germany}

\author{Tomohiro Matsui}
\affiliation{Institute of Applied Physics, University of Hamburg,
Jungiusstrasse 11 20355 Hamburg, Germany}

\author{Christian Heyn}
\affiliation{Institute of Applied Physics, University of Hamburg,
Jungiusstrasse 11 20355 Hamburg, Germany}

\author{Wolfgang Hansen}
\affiliation{Institute of Applied Physics, University of Hamburg,
Jungiusstrasse 11 20355 Hamburg, Germany}

\author{Roland Wiesendanger}
\affiliation{Institute of Applied Physics, University of Hamburg,
Jungiusstrasse 11 20355 Hamburg, Germany}

\date{\today}

\begin{abstract}
We show both theoretically and experimentally that 
scanning tunneling spectroscopy (STS) images
of semiconductor quantum dots may display clear signatures 
of electron-electron correlation. We apply many-body tunneling
theory to a realistic model which fully takes into account
correlation effects and dot anisotropy. Comparing measured STS images 
of freestanding InAs quantum dots with those calculated by
the full configuration interaction method, we explain the
wave function sequence in terms of images of one- and two-electron
states. The STS map corresponding to double charging is
significantly distorted by electron correlation with 
respect to the non-interacting case.

\end{abstract}


\maketitle

\section{Introduction}

Scanning tunneling microscopy (STM) and spectroscopy (STS) is a key 
tool in nanoscience, allowing for both manipulation of nano-objects
and access to their energy spectrum and wave function (WF). 
The class of systems under study is broad and loosely
defined, ranging from nanotubes to quantum dots (QDs) and molecules.
Their common features are the possibility of achieving relatively good 
electrical insulation, the discreteness of their 
energy spectrum, and the manifestation of Coulomb blockade phenomena 
at low temperatures. In these few-body systems electron-electron interaction 
may play a major role, as it is immediately apparent e.g.~from 
single-electron charging experiments.\cite{Grabert92} 

Some of us recently suggested that the WF imaging technique of STS
could be a sensitive and direct probe of electron 
correlation.\cite{Rontani05b,Rontani06b,RontaniBAbis,Maruccio06}
Despite recent experimental evidence that few-body semiconductor QDs 
are strongly affected by electron correlation, as seen both in 
inelastic light scattering\cite{Garcia05} and high 
source-drain\cite{Korkusinski04} spectroscopies, all QD WF
images obtained so far, both in 
real\cite{Grandidier00,Millo01,Maltezopoulos03} and 
reciprocal\cite{Vdovin00,Patane02,Wibbelhoff05,Kailuweit06} space, were 
basically interpreted in terms of independent-electron WFs.
In this paper we apply many-body tunneling theory to a realistic
model which fully takes into account the combined effect of dot anisotropy
and Coulomb interaction and predict that STS maps can be strongly
distorted by correlation effects. 
These calculations are validated
by a recent low-temperature-STS experiment performed by a few of 
us on few-electron InAs QDs at the University of Hamburg, 
which was reported in Ref.~\onlinecite{Maruccio06} and that we
here review providing further details and discussion
of the theoretical model.

While in this paper we focus exclusively
on semiconductor QDs, we stress that the main ideas
regarding how electron correlation affects STS could in principle be applied
also to short carbon nanotubes as well as single molecules.

The plan of the paper is as follows:
After a review of conceptual frameworks used to understand STS
(Sec.~\ref{s:concept}), we illustrate our full configuration
interaction (FCI) method of
calculation of STS images (Sec.~\ref{s:FCI})
and theoretically focus on the double charging of a QD 
in various realistic regimes (Sec.~\ref{s:predictions}).
We discuss the evidence of correlation effects
provided by the Hamburg experiment (Sec.~\ref{s:Hamburg}),
and eventually reconsider our findings in the 
Conclusions (Sec.~\ref{s:conclusions}).

\section{Theory of the STS in the presence of electron 
correlation}\label{s:concept}

In order to understand the origin of correlation effects in STS, it
is useful to critically reconsider the basic picture of STS performed
in the presence of Coulomb blockade. Figure 1 displays 
schematically the typical energy landscape along the tunneling 
direction seen by an electron flowing from the STM tip to a drain
lead (backgate in Fig.~1) through the QD. The QD is
well isolated from tip and backgate by two tunneling barriers
(the vacuum and an insulating layer, respectively), 
so the width of its discrete energy levels is negligible.
At low temperature transport occurs mainly due to resonant tunneling
through the QD chemical potentials $\mu(N) = E_0(N) - E_0(N-1)$,
which are depicted as segments in the QD region of Fig.~1 
[$N$ is the number of electrons filling in the QD at equilibrium
and $E_0(N)$ is its ground-state energy].\cite{notacharging} 
If the tip-backgate voltage $V$ is small, current may flow or
not (Coulomb blockade)
depending if some value of $N$ exists such that $\mu(N)$
enters the transport energy window.
By increasing $V$ one is able to widen the
transport window, causing a step in the current $I$ to occurr
(a peak in the differential conductance $dI/dV$ in the inset of 
Fig.~1) each time the chemical potential
$\mu(N')$ for a new electron number $N'$ enters the window.

In the standard STS mean-field theory,\cite{Tersoff85} 
the chemical potentials $\mu(N)$'s are simply Hartree-Fock or 
Kohn-Sham single-particle (SP) self-consistent levels,
split in energy by Coulomb and exchange interactions in addition 
to the effect of quantum confinement. The energy- and space-resolved
local density of SP states, $n(\varepsilon,\bm{r})$,
is just given by a sum over the occupied SP orbitals $\psi_{\alpha}
(\bm{r})$, $n(\varepsilon,\bm{r})=\sum_{\alpha}\left|\psi_{\alpha}
(\bm{r})\right|^2\delta(\varepsilon-\varepsilon_{\alpha})$,
where $\varepsilon_{\alpha}$ is the energy of the $\alpha$th
SP level.
A general result of many-body time-dependent perturbation theory,
valid to first order in the tunneling matrix element,\cite{Bardeen61}
is that $dI/dV$ is proportional to the local electron density. 
\begin{figure}
  \includegraphics[width=0.7\columnwidth]{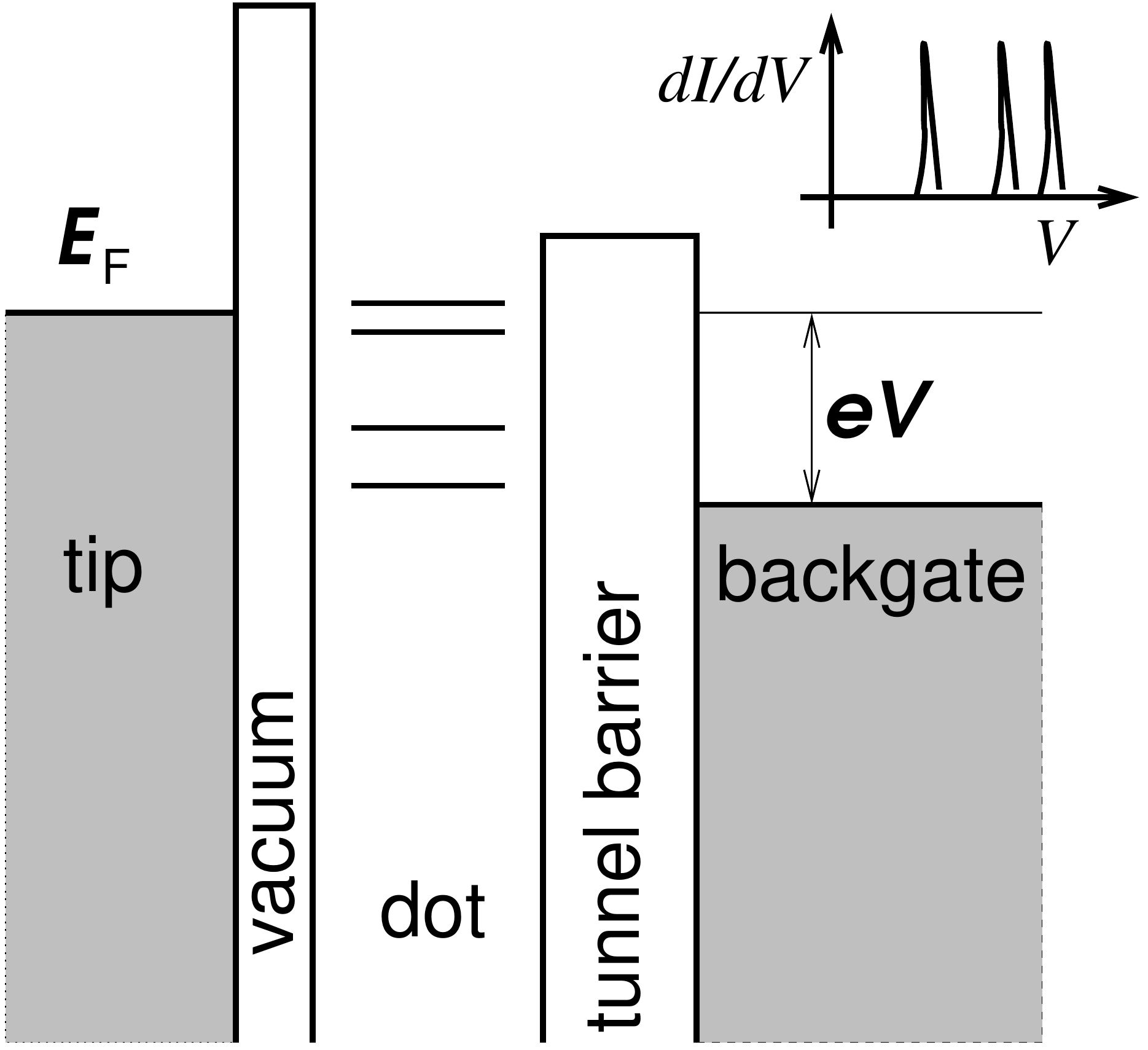}
  \caption{
Simplified energy landscape along the direction of tunneling
through a quantum dot for a typical scanning tunneling
spectroscopy measurement. $E_{\text{F}}$ is the Fermi energy
of the STM tip. Upper right inset: Corresponding
idealized $dI/dV$ vs. $V$ plot (ignoring all broadening mechanisms
but resonant tunneling).
}
\label{fig1}
\end{figure}
Therefore, in the framework of mean-field theory,
the local value of the differential conductance, $dI/dV(E_{\text{F}},\bm{r})$,
is given by a sum over the square moduli of SP orbitals whose
energies fall in the transport window $\delta\varepsilon$
around the Fermi energy $E_{\text{F}}$,
$dI/dV(E_{\text{F}},\bm{r})\propto \sum_{\delta\varepsilon} 
\left|\psi_{\alpha}
(\bm{r})\right|^2$. If the energy resolution is sufficient, 
the space-resolved map of $dI/dV(E_{\text{F}},\bm{r})$ provides the image 
of the square modulus of a {\em single} SP orbital, namely the 
highest-energy occupied QD orbital, akin to the HOMO in molecules. 
Besides, the STS energy-scan mode exhausts the {\em Aufbau} 
filling sequence of the lowest-energy SP levels of the dot.

Since in general the QD energy spectrum strongly depends 
on electron-electron interaction and consequently
on $N$,\cite{ReimannRMP} the above mean-field theory suffers 
a few serious drawbacks: (i) Each tunneling event involves the
transition between QD ground states with $N$ and $N+1$ electrons,
therefore it is unclear which value of $N$ should be used in 
the $N$-dependent computation of self-consistent SP orbitals 
$\psi_{\alpha}$.\cite{notaextended}
(ii) The interacting $N$-electron ground state $\left|\Psi_N\right>$
is in general a linear superposition of many electronic configurations
(Slater determinants), $\left|\Phi_i^N\right>$, with expansion coefficients
$c_i^N$, $\left|\Psi_N\right> = \sum_i c_i^N \left|\Phi_i^N\right>$,
while mean-field theory approximates $\left|\Psi_N\right>$
with a single Slater determinant, 
$\left|\Phi_1^N\right>$ ($c_i^N=\delta_{i1}$):
this approximation turns out to be poor in strongly
correlated regimes.\cite{Rontani06}

In order to circumvent the above difficulties we recall
from many-body tunneling theory that the differential
conductance is proportional to the {\em interacting}
local density of states,\cite{Feuchtwang74}
\begin{equation}
dI/dV(E_{\text{F}},\bm{r})\propto -\frac{1}{\hbar\pi}\text{Im}
\mathcal{G}(\bm{r},\bm{r};E_{\text{F}}),
\label{eq:Green}
\end{equation}
where $\mathcal{G}(\bm{r},\bm{r};E_{\text{F}})$ is the energy-
and space-resolved exact zero-temperature retarded Green's function
(or one-electron propagator).\cite{Rontani06c}
The quantity on the r.h.s. of Eq.~(\ref{eq:Green}) (also known as
spectral density) may be regarded as the squared modulus of a 
{\em quasi-particle} WF:\cite{Rontani05b}
\begin{equation}
\left|\varphi_{\text{QD}}(\bm{r})\right|^2 =
-\frac{1}{\hbar\pi}\text{Im}
\mathcal{G}(\bm{r},\bm{r};E_{\text{F}})   .
\end{equation}
The quasi-particle WF $\varphi_{\text{QD}}(\bm{r})$
is the natural generalization of the SP WF to strongly
correlated regimes. In the non-interacting limit
$\varphi_{\text{QD}}(\bm{r})\rightarrow \psi(\bm{r})$, as we show below.
In practice, the calculation of $\varphi_{\text{QD}}(\bm{r})$  
requires the knowledge of the (configuration interaction)
expansion coefficients $c_j^N$ and $c_i^{N-1}$
of the interacting ground-states $\left|\Psi_N\right>$ and
$\left|\Psi_{N-1}\right>$ with both $N$ and $N-1$ electrons,
respectively, according to the formula $\varphi_{\text{QD}}(\bm{r})
=\sum_{i,j}c_i^{N-1*}c_j^N\psi_{\alpha(i,j)}(\bm{r})$,
where $\alpha(i,j)$ is a SP quantum index depending on both
Slater determinants
$\left|\Phi^{N-1}_i\right>$ and $\left|\Phi^N_j\right>$.
Distortion effects of $\varphi_{\text{QD}}(\bm{r})$ with 
respect to $\psi(\bm{r})$ are due to the quantum interference 
among different SP orbitals $\psi_{\alpha(i,j)}(\bm{r})$.
In the non-interacting limit, both $N$- and $(N-1)$-electron
ground states are single Slater determinants, $c_i^{N-1}=\delta_{i1}$,
$c_j^N=\delta_{j1}$, and therefore $\varphi_{\text{QD}}(\bm{r})$
reduces to the simple SP orbital $\psi_{\alpha(1,1)}(\bm{r})$.

\section{Predicting STS maps from full configuration 
interaction}\label{s:FCI}

We consider a two-dimensional QD with parabolic lateral
confinement, which is a commonly accepted approximation\cite{ReimannRMP}
in the framework of the envelope-function description
of SP states:
\begin{equation}
H_{0}(i)\,=\,\frac{\bm{p}_i^2}{2m^{*}}
+\frac{1}{2}m^*\!\!\left(\omega_x^2x_i^2 + \omega_y^2y_i^2\right).
\label{eq:HSP}
\end{equation}
In the SP Hamiltonian for the $i$th electron of Eq.~(\ref{eq:HSP})
the lateral confinement is different in the $x$ and $y$ directions
(with corresponding confinement frquencies $\omega_x$ and
$\omega_y$, respectively): Such an elliptical confinement mimics 
the combined effects of geometrical deviations from
perfect circularity and/or atomistic effects\cite{Zunger05} due to
strain, piezolectric fields, interface matching, which 
lower the symmetry point-group from $D_{\infty h}$ (circular case)
to $C_{2v}$ [the symmetry point-group of the Hamiltonian
(\ref{eq:HSP}), $D_{2h}$, is actually slightly larger than 
the $C_{2v}$ group]. In addition, we include into
our model the full Coulomb interaction among the $N$ electrons
populating the dot:
\begin{equation}
H = \sum_{i}^{N}H_{0}(i)+\frac{1}{2}\sum_{i\neq j}\frac{e^{2}}
{\kappa|\bm{r}_i-\bm{r}_j|}.
\label{eq:HI}
\end{equation}
In Eqs.~(\ref{eq:HSP}-\ref{eq:HI}) $e$ and $\kappa$ are respectively 
the electron charge and static relative dielectric constant of the host
semiconductor, 
$\bm{p}$ is the canonically conjugated momentum of position
$\bm{r}\equiv (x,y)$, $m^*$ is the electron effective mass.

We solve numerically the few-body problem of Eq.~(\ref{eq:HI})
for the ground states of $N$ and $N-1$ electrons 
by means of the FCI method, that we successfully applied in
predicting QD transport\cite{Ota05} and Raman\cite{Garcia05} spectra
(for full details on our FCI method, its performances,
and ranges of applicability, see Ref.~\onlinecite{Rontani06}).
Briefly, we expand the $N$-electron interacting ground state
$\left|\Psi_N\right>$ on the basis of the Slater determinants 
$\left|\Phi_i^N\right>$ obtained by filling in with $N$ electrons
in all possible ways a subset of the SP orbitals,
eigenstates of Hamiltonian (\ref{eq:HSP}).
On the Slater-determinant basis, the Hamiltonian (\ref{eq:HI}) 
is a large, sparse
matrix that we diagonalize by means of a parallel state-of-the-art
Lanczos code. The expansion coefficients $c_i^N$, which are
the output of the FCI calculation, are eventually used to
build the quasi-particle WF $\varphi_{\text{QD}}(\bm{r})$.

\section{Quasi-particle images in realistic quantum dots: 
Theory}\label{s:predictions}

We here consider quasi-particle WF images for realistic QDs
predicted by the theory of
Secs.~\ref{s:concept}-\ref{s:FCI},
focusing on the simplest case where electron
correlation becomes relevant, namely the tunneling transition
$N=1\rightarrow N=2$. In fact, the image of the first 
charging process $N=0\rightarrow N=1$ is simply given by the 
lowest-energy SP orbital.

We first investigate the case of a circular 
QD by monitoring the evolution of the square modulus of 
$\varphi_{\text{QD}}(\bm{r})$ as a function of the lateral
confinement energy $\hbar\omega_0=\hbar\omega_x=\hbar\omega_y$.
On a general basis, we expect that correlation effects are
negligible in the non-interacting limit $\hbar\omega_0\rightarrow
\infty$ while they dominate in the opposite limit
$\hbar\omega_0\rightarrow 0$. In fact, the SP [Eq.~(\ref{eq:HSP})]
and the Coulomb [Eq.~(\ref{eq:HI})] terms of the interacting
Hamiltonian scale differently with $\hbar\omega_0$, the former as
$\sim\hbar\omega_0$ and the latter as 
$\sim\sqrt{\hbar\omega_0}$ (Ref.~\onlinecite{Rontani05b}).
When $\hbar\omega_0\rightarrow \infty$ the SP term dominates
with respect to the Coulomb term, and the ground state is 
essentially a single Slater determinant, while interaction effects
can be regarded as a perturbation: in this limit mean-field theory
is expected to give correct predicitons. In the opposite limit,
instead, the Coulomb term very effectively mixes many Slater 
determinants, and significant distortions of $\varphi_{\text{QD}}(\bm{r})$
with respect to the non-interacting case are expected.

\begin{figure}
  \includegraphics[width=1.0\columnwidth]{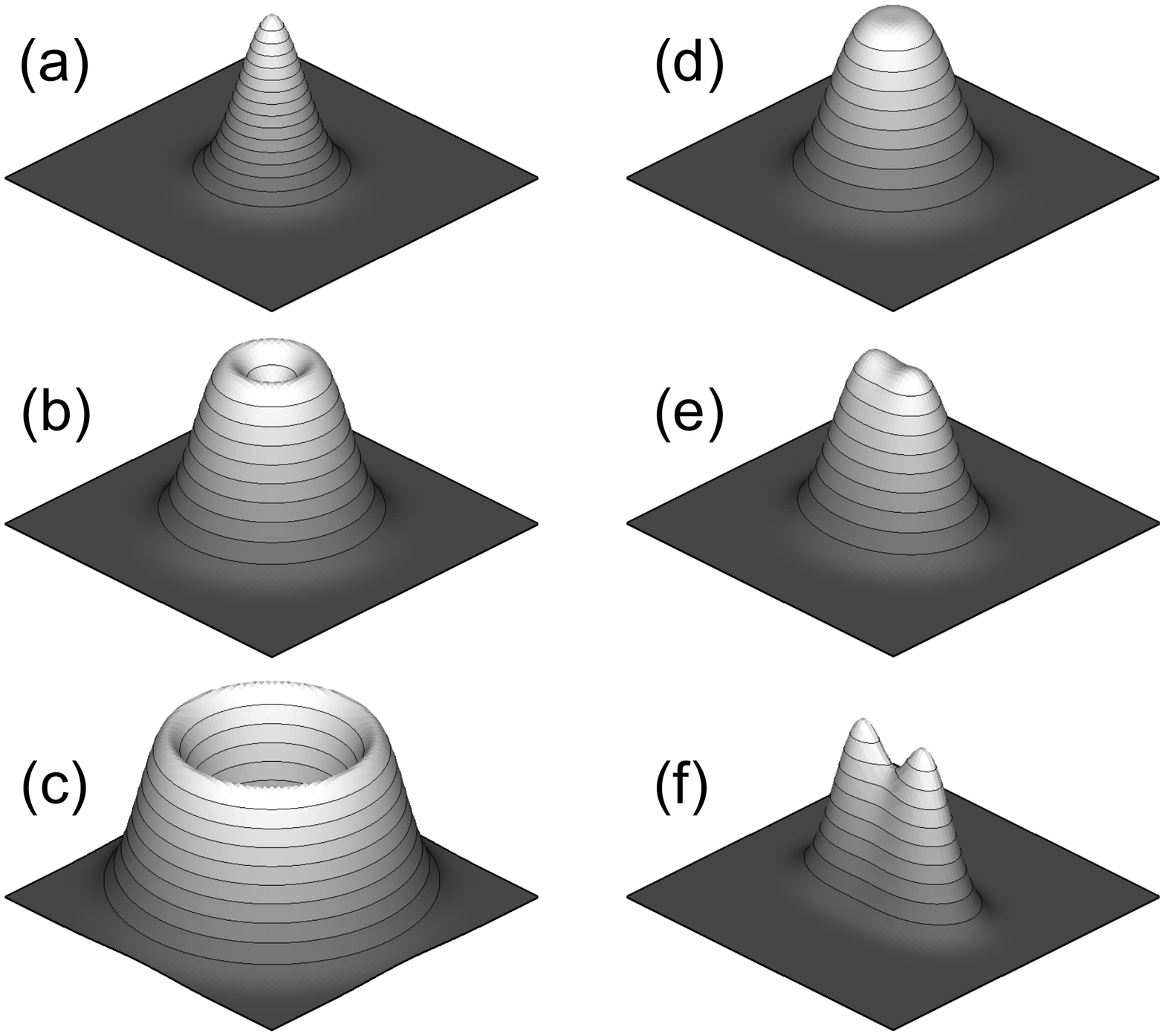}
  \caption{Calculated STS maps for the ground state $\rightarrow$
ground state transition $N=1\rightarrow N=2$, as a function of
lateral confinement energy $\hbar\omega_0$ and dot anisotropy.
(a-c) Dependence of the STS image on the
lateral confinement energy $\hbar\omega_0$ in the case of a
circular dot. Plots (a-c) correspond to $\hbar\omega_0=$
$+\infty$,
0.5, 0.01 meV, respectively, with $\omega_0=\omega_x=\omega_y$.
(d-e) Dependence of the STS image on the
dot anisotropy. Plots (d-e) correspond to $\omega_x/\omega_y=$ 1,
1.1, 1.5, respectively, with $\hbar\omega_y=$ 2 meV. The lateral extension
of all plots is 4 $\times$ 4 units of the characteristic lateral
extension of the harmonic oscillator $\ell_{\text{QD}}
=[\hbar/(m^*\omega_y)]^{1/2}$,
and GaAs parameter are used throughout the paper.
Plot heights are renormalized arbitrarily. The absolute
norms of STS maps are given in Table \ref{t:tab1}.
}
\label{fig2}
\end{figure}

Figures 2(a-c) display the maps of 
$\left|\varphi_{\text{QD}}(\bm{r}) \right|^2$ in the $(x,y)$
plane for decreasing values of $\hbar\omega_0$ [from top
to bottom, $\hbar\omega_0\rightarrow +\infty$ (a), 
$\hbar\omega_0 =$ 0.5 meV (b), $\hbar\omega_0 =$ 0.01 meV (c)].
In the non-interacting case [Fig.~2(a)], the STS map
is just a replica of the SP 1$s$ orbital, eigenstate of the
2D harmonic oscillator (a 2D gaussian), corresponding to the injection
of a second electron into the same 1$s$ level occupied by the first 
electron with opposite spin, according to Pauli
exclusion principle. As $\hbar\omega_0$ is decreased, we see that
a significant portion of weight is moved from the image center
into an outer ring [Fig.~2(b)]. In a very shallow
dot [Fig.~2(c)], the dot center is completely emptied
and the STS image now looks like a donut. The latter case
of strong interaction shows that the STS image can be
very different from that expected by a naive application
of the Aufbau principle, due to the hybridization of higher-energy
SP orbitals with the 1$s$ WF. 

In Fig.~2 the length unit is the characteristic lateral
extension of the harmonic oscillator, $\ell_{\text{QD}}
=[\hbar/(m^*\omega_y)]^{1/2}$, which depends on $\hbar\omega_y$,
and $\varphi_{\text{QD}}(\bm{r})$ is arbitrarily normalized.
Another effect of correlation is the loss of 
absolute weight of the quasi-particle WF as $\hbar\omega_0$ is 
decreased.\cite{Rontani05b}
\begin{table}
\caption{\label{t:tab1}Norm of the quasi-particle,
$\int d\bm{r} \left|\varphi_{\text{QD}}(\bm{r})\right|^2$,
for different ground-state $\rightarrow$ ground-state
tunneling transitions $N=1\rightarrow N=2$. The weight
ranges between 0 and 1.}
\begin{ruledtabular}
\begin{tabular}{cccc}
Case\footnote{The case refers to the calculated STS image
displayed in Fig.~\ref{fig2}.}
& Quasi-particle weight & Case & Quasi-particle weight \\
\hline
(a) & 1.00  & (d) & 0.794 \\
(b) & 0.633 & (e) & 0.793 \\
(c) & 0.109 & (f) & 0.720 \\
\end{tabular}
\end{ruledtabular}
\end{table}
Table \ref{t:tab1} shows the
norm of $\varphi_{\text{QD}}(\bm{r})$
for the cases displayed in Fig.~2.
We see that the reduction of the confinement energy 
$\hbar\omega_0$ in a circular
QD is associated to a dramatic weight loss. This trend may
be seen as a signature of the Wigner crystallization of 
the $(N-1)$- and $N$-electron ground states which manifests 
itself as an increased ``rigidity'' of the states opposing 
electron injection.

We now switch to consider the effect of dot anisotropy,
since the dependence of STS images on dot ellipticity
 was found to be a major issue in the Hamburg experiment
(cf.~Sec.~\ref{s:Hamburg} below).\cite{Maruccio06}
In Fig.~2(d) we start showing the predicted STS map 
for a circular QD with $\hbar\omega_0=$ 2 meV, and then 
modify its ellipticity by increasing the lateral frequency
ratio $\omega_x/\omega_y$, going from $\omega_x/\omega_y=$
1.1 [Fig.~2(e)]
up to $\omega_x/\omega_y=$ 1.5 [Fig.~2(f)]. 
By doing so, as we move downwards in
the right column of Fig.~2, we expect
to ``squeeze'' the quasi-particle WF along the $x$ direction,
which is indeed observed for both single and double charging 
processes.  However, in the present double charging case,
we also see an unexpected effect, i.e.~the quasi-particle
WF develops two peaks along the $y$-axis.
This surprising distortion is due to the destructive
interference between 1$s$ and 1$d$ states of the
harmonic oscillator along $y$ (belonging to the
same representation $A$ of the $D_{2h}$ group), which is
a correlation effect. This can also be seen as a manifestation
of the general statement that the importance of correlation 
increases as the system dimensionality is reduced (from 2D to 1D).

To conclude this section, we mention that another issue
relevant to STS images is the type of dielectric environment 
felt by QD electrons.\cite{Maruccio06} Specifically, the dielectric
mismatch between vacuum and InAs in the Hamburg experiment affects
both the SP confinement potential (self-polarization effect)
and the electron-electron interaction (interaction with
surface image charges of like sign), therefore changing
the relative importance of their effect on
the ground state. Qualitatively, we expect that changing the
dielectric environment causes effects similar to those obtained 
by modifying $\hbar\omega_0$.

\section{STS imaging of few-electron MBE-grown quantum 
dots: Experiment}\label{s:Hamburg}

We studied strain-induced InAs QDs grown on n-doped GaAs(001)
substrates by molecular beam epitaxy (MBE).\cite{Maltezopoulos03}
An undoped tunneling barrier 5 nm thick insulates the 
freestanding QDs from a n-doped GaAs buffer layer 200 nm
thick, acting as a backgate for the current flowing from
the STM tip through the QDs. The samples were transferred
from the MBE into the STM chamber in ultra-high vacuum 
without being exposed to the air by means of a mobile 
system, in ordered to avoid contamination,
and STM was operated at a background pressure smaller than $10^{-10}$ mbar 
and at $T=$ 6 K with maximum energy resolution of $\delta\varepsilon=$
2 meV.\cite{Wittneven97} STM images were taken in constant-current
mode, with a typical sample bias in the range of 2-4 V
and a tunneling current of 20-40 pA. A lock-in 
technique (modulation voltage $V_{\text{mod}}$ in the range
of 5-20 mV) was used to record $dI/dV(E,\bm{r})$
and WF mapping was carried out over a specified area by stabilizing
the tip-surface distance in each point $\bm{r}$ at voltage
$V_{\text{stab}}$ and current $I_{\text{stab}}$, switching 
off the feedback and recording a $dI/dV$ curve from $V_{\text{start}}$
to $V_{\text{end}}$ ($V_{\text{start}}\leq 
V_{\text{stab}}$).\cite{Maltezopoulos03}
As a result, WF mapping produces a 3D array of $dI/dV$ data,
which allows obtaining spatially resolved $dI/dV(E,\bm{r})$
images at different values of sample voltages. 

\begin{figure}
  \includegraphics[width=0.25\columnwidth,angle=90]{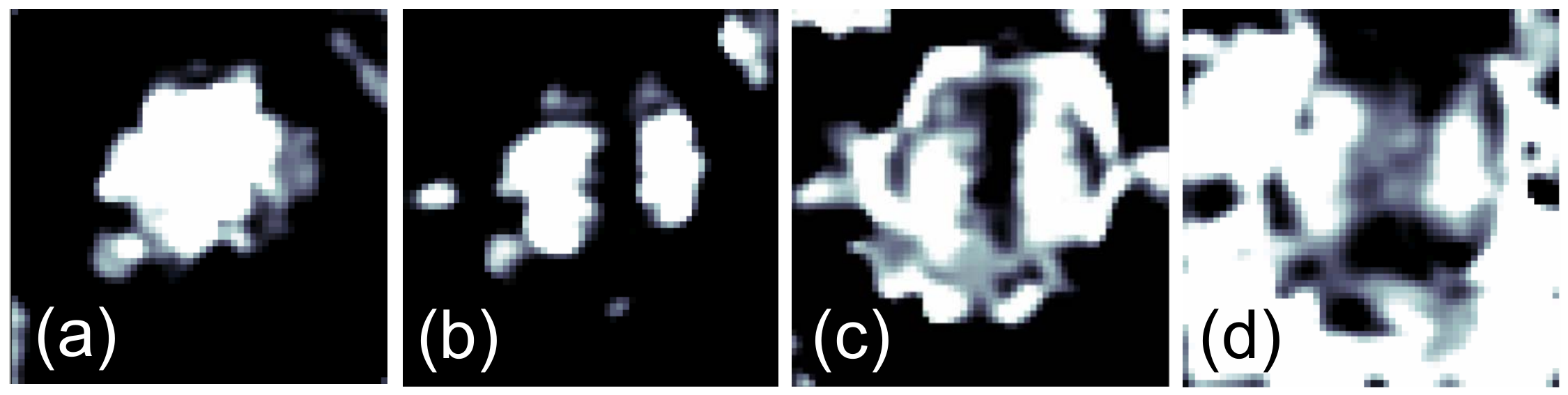}
  \caption{$dI/dV(E,\bm{r})$ spatial
maps recorded at the voltages 840 (map a),
1040 (b), 1140 (c), and 1350 (d) mV, respectively, for
a representative QD.
The size of all maps is 30 $\times$ 30 nm.
White (black) colour stands for high (low) values of $dI/dV$.
$I_{\text{stab}}=$ 100 pA, $V_{\text{stab}}=$ 1500 mV,
$V_{\text{mod}}=$ 4 mV.
}
\label{fig3}
\end{figure}

In Fig.~3 we show $dI/dV(E,\bm{r})$ maps of a representative QD
measured at fixed voltages corresponding to  
four clearly localized QD WFs, labeled (a), (b), (c), and (d), respectively,
in order of increasing energy.\cite{Maruccio06}
We observed a pronounced shape anisotropy for all dots,
which have a pyramidal shape with well-defined facets,
consistently with previous findings.\cite{Maltezopoulos03,Marquez01}
This is evident also from the elongation of images (b), (c), and (d)
of Fig.~3 along the [1\underbar{1}0] direction.
In detail, WF maps of  Fig.~3 show the following approximate
symmetries, going from low to high-energy: 
$s$-like for (a), $p$-like for (b) and (c), and possibly (d). The (d) image
is somehow blurred, likely because states of the wetting layer
overlapping in energy with (d)
significantly contribute to the spectral density.
See Ref.~\onlinecite{Maruccio06} for further discussion.

As expected for the square moduli of the two lowest-energy SP 
1$s$ and 2$p$ orbitals, images (a) and (b) exhibit 
a roughly circular symmetric intensity distribution and 
elongation along the [1\underbar{1}0] direction with a node in the
center, respectively (Fig.~3). Unexpectedly, image (c) shows
again a $p$-like symmetry in the [1\underbar{1}0] direction, 
as before, instead of [110] as expected for the second 2$p$ 
orbital.\cite{Maltezopoulos03} As a consequence, it is not 
possible to explain the WF sequence [and map (c) in particular]
in terms of SP orbitals, since in this case we would expect the 
appearance of either a single 2$p$ or two 2$p$ states
elongated in the [1\underbar{1}0] and [110] directions, respectively.
On the other hand, from a mean-field point of view the charging  
of the same 2$p$ orbital with a second electron can be excluded
since a replica is not observed for the 1$s$ orbital.

On this basis, we believe that the understanding of
the data of Fig.~3 must rely on the theory
developed in Sec.~\ref{s:predictions}, taking into account
the combined effect of electron correlation, dot
anisotropy, and dielectric mismatch. Specifically,
we assign maps (a) and (b) to the tunneling events
$N=0\rightarrow N=1$ corresponding to resonances through 
1$s$ and 2$p$ SP states,
while we associate image (c) of Fig.~3
to the ground state $\rightarrow$
ground state charging process $N=1\rightarrow N=2$, showing
features similar to those predicted in Figs.~2(e-f). 
Further support for this interpretation is reported in 
Ref.~\onlinecite{Maruccio06}.
 
\section{Conclusions}\label{s:conclusions}

In this paper we have focused on QDs where electron-electron
interaction may be relevant, showing that the STS technique
is sensitive to
quasi-particle WFs and that their images can be greatly
affected by electron correlation. On the basis of FCI
calculations of STS maps which fully take into account correlation
effects and dot anisotropy, we have been able to understand
measured WF images of freestanding self-assembled
InAs quantum dots. We identified
ground- and excited-state WFs corresponding to the injection
of a first and a second electron into the QD. Correlation
effects are found to distort the WF corresponding to double charging 
in an essential way. We believe our findings may be relevant
to a broader class of experiments, including STS on 
carbon nanotubes and single molecules.

\section*{Acknowledgments}

This work is supported by the INFM-CINECA Supercomputing Project 2006,
the MIUR-FIRB Project RBIN04EY74, the EU-network project Nanospectra,
and the Deutsche Forschungsgemeinschaft (SFB508, TP A6 and B7).

%


\begin{thebibliography}{26}
\bibitem[{not({\natexlab{a}})}]{notacharging}
\bibinfo{note}{For the sake of simplicity, we here ignore contribution of
  excited states to transport as well as higher-order and non-equilibrium
  effects.}
\bibitem[{not({\natexlab{b}})}]{notaextended}
\bibinfo{note}{Note that the question is irrelevant for extended systems where
  self-consistent SP orbitals do not depend significantly on $N$.}
\end{thebibliography}

\end{document}